\documentclass[fleqn,12pt,twoside]{article}
\usepackage{espcrc1}

\usepackage{graphicx}
\usepackage[figuresright]{rotating}

\usepackage{color}
\usepackage{amssymb}

\newcommand{\AmS}{{\protect\the\textfont2
  A\kern-.1667em\lower.5ex\hbox{M}\kern-.125emS}}
\definecolor{vert}{rgb}{0,0.5,0} 

\title{Nucleon longitudinal spin structure at COMPASS~\footnote{Prepared for 10th International Baryons Conference (Baryons 2004), Palaiseau, France, 25-29 Oct 2004.}}

\author{Y.Bedfer\address{DAPNIA/SPhN, CEA Saclay	 \\
        F91191 Gif/Yvette, France} for the COMPASS collaboration}
       
\begin{document}

\maketitle

\begin{abstract}
The nucleon spin structure is studied at COMPASS via the
measurement of the spin dependent properties of
deep inelastic muon-deuteron scattering. Both inclusive and
semi-inclusive observables are investigated, with the emphasis
put on the latter. This allows to access, in particular,
 the gluon polarization ($\Delta G/G$)
and the flavor decomposition of
the quark helicity distributions.

Results from measurements with longitudinal target polarization are presented, as well as uncertainties expected from the first 3 years of data taking.
\end{abstract}

\section{INTRODUCTION}
COMPASS~\cite{spectro} is a fixed-target
experiment installed on the CERN SPS M2 beam line. One of its goals is the unraveling of the spin structure of the nucleon. In
this respect, it attempts to answer questions left open
by the measurements of the double spin asymmetry in
inclusive deep inelastic lepton-nucleon scattering $A_1$.

  First, the gluon distribution $\Delta G$ remains largely undetermined. QCD fits of
present world data can only but poorly constrain it from the
$Q^2$ dependence of $A_1$, for lack
of lever arm in $Q^2$.

  Secondly, inclusive data alone do not allow to separate quark and
anti-quark distributions.

 In COMPASS, additional information is brought by
semi-inclusive measurements. These are performed with transverse as well
as longitudinal target spin orientations, which allows  to
investigate also the transversity distribution, {\it cf.}~\cite{transverse}.

 Finally, in the inclusive measurement of $A_1$, COMPASS
 reaches unprecedented precision in the low Bjorken $x$
domain, which will bring improvement in the knowledge of the integral of
the $g_1$ structure function.

\section{EXPERIMENTAL ESSENTIALS}
 The COMPASS spectrometer is described in~\cite{spectro}. 
Its experimental setup was designed to allow a precise determination of
asymmetries. Thanks to its simultaneous measurement of both orientations of
target spin in 2 oppositely polarized target cells, $u$ and $d$, and
to a frequent reversal of spin orientations, the acceptance and incident
muon flux cancel out in the formula for the cross-section asymmetry $A_{{\mu}N}$:
\begin{equation}
A_{{\mu}N} = \frac{1}{2}\quad \frac{1}{P_b \times P_tf}
  \left(\frac{N_u^{\Uparrow\uparrow}-N_d^{\Uparrow\downarrow}}{N_u^{\Uparrow\uparrow}+N_d^{\Uparrow\downarrow}}+
  \frac{N_d^{\Uparrow\uparrow}-N_u^{\Uparrow\downarrow}}{N_d^{\Uparrow\uparrow}+N_u^{\Uparrow\downarrow}}\right)
\label{AmuN}
\end{equation}
where $N$ is the counting rate, $\Uparrow\uparrow$ and $\Uparrow\downarrow$
denote the 2 spin configurations, $P_b$ is the beam polarization ($\sim$76\%),
$P_t$ and $f$ are the $^6LiD$ target polarization ($\sim$50\%) and,
 process dependent, dilution factor ($\sim$40\%).

 The experiment has accumulated $\sim$3~$fb^{-1}$ of data in its 3 years of running
from 2002 to 2004. About half of these data has been analyzed so far.

\section{\boldmath$\Delta G/G$}
\begin{figure}[h]
  \begin{minipage}[b]{0.44\linewidth}
    \includegraphics*[width=0.99\linewidth]{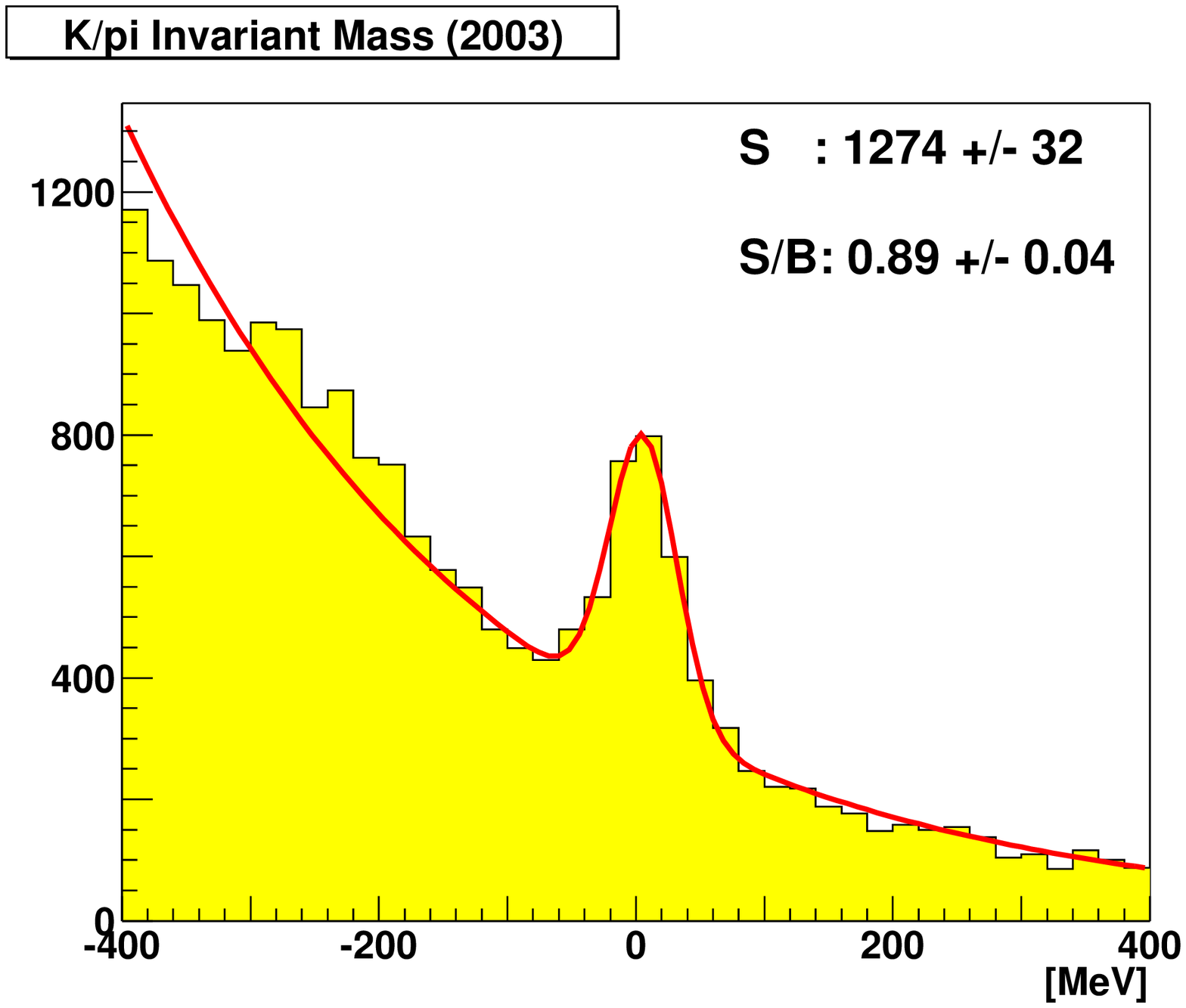}
    \vspace*{1cm}
  \end{minipage}
\includegraphics*[width=0.54\linewidth]{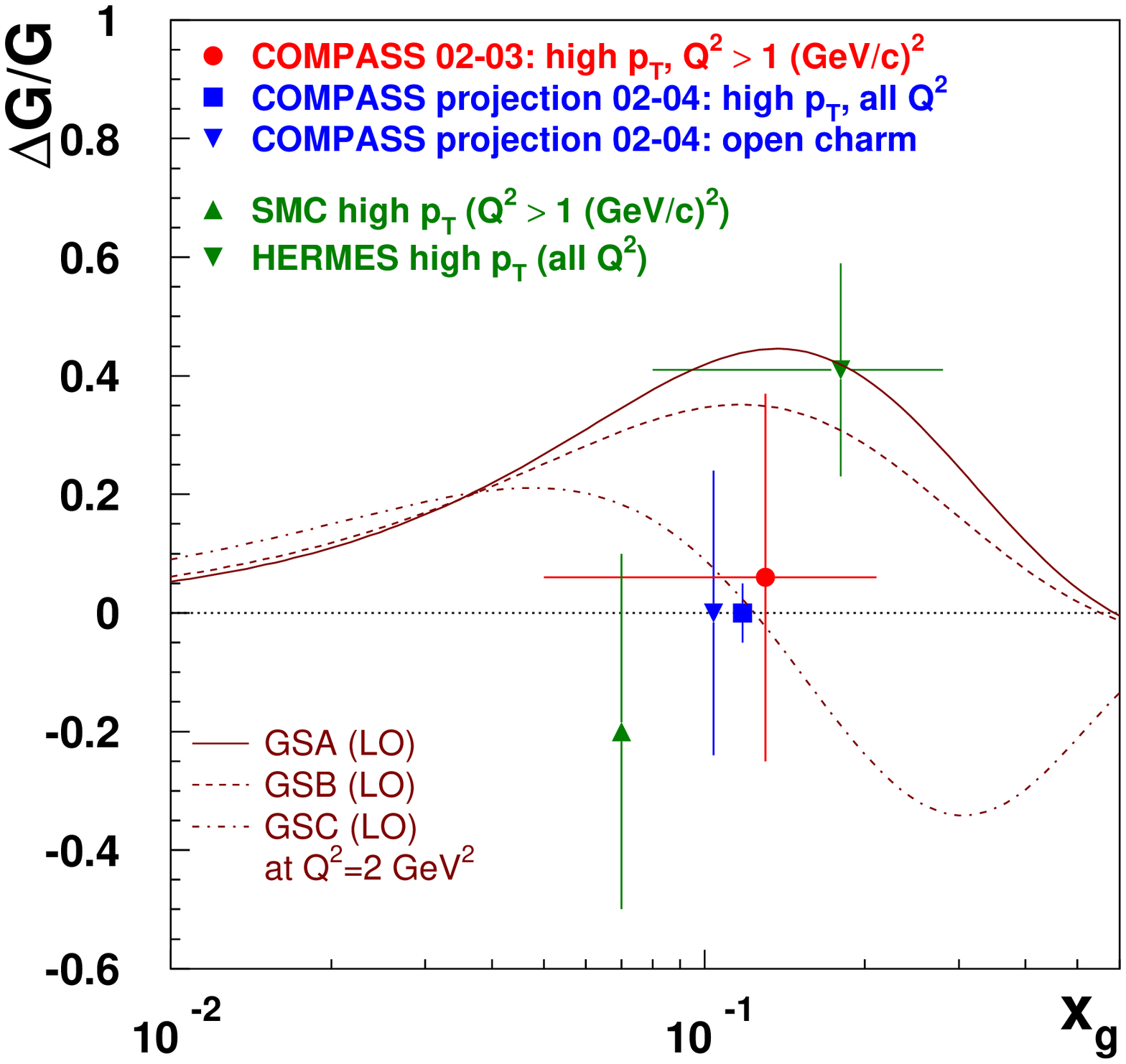}
\vspace*{-1.5cm}
\caption{Left: $D^o$ peak in the $K\pi$ invariant mass distribution of $D^*$-tagged events for 2003 ({\it i.e.} $\sim$1/4 of total) data.
	Right: $\Delta G/G$: Results from
	 COMPASS high $p_T$ $Q^2>1$ (2002-2003 data, {\it i.e.} $\sim$1/2 of total) ~\cite{pt2}, compared with results from
	HERMES~(all $Q^2$)~\cite{hermes}
	and SMC~($Q^2>1$)~\cite{smc},
	 and projected error for
	COMPASS high $p_T$ all $Q^2$ and open charm channels from all of
2002-2004 data. The curves show the
parameterizations of $\Delta G/G$ {\it vs.} the momentum fraction carried by the gluon, $x_g$, from~\cite{GS}.}
\label{DG}
\end{figure}

 In COMPASS, we access the gluon distribution via the photon-gluon fusion $PGF$
process, whereby a virtual photon couples to a gluon via a $q\bar{q}$ pair.
We consider two
different selections of the $PGF$: open charm and high $p_T$ hadron
production. Both have been successfully used to directly measure the
unpolarized gluon distribution at the HERA collider experiments (with the
difference that high $p_T$ refers to the production of jets there).
For both, hard interactions are selected irrespective of
$Q^2$, by, respectively, the charm mass and
the $p_T$ cut. But the 2 selections differ largely in terms of statistics (higher for high $p_T$)
and background. In the high $p_T$ case, not only do physics background processes
contribute  significantly
but their impact can only
be evaluated by a Monte Carlo modelization of the experiment. The intricacy of
the problem becomes larger at low $Q^2$, where processes involving resolved
photons dominate. Therefore, two sub-cases are further considered, {\it viz.}
$Q^2>1$ and all $Q^2$. A determination of $\Delta G/G$ has been obtained for
the sub-sample with $Q^2>1$ and for the much bigger sample with all $Q^2$
only the projection for the precision expected
from the 2002-2004 data is so far available. The 2 values
are plotted on Fig.~\ref{DG} and discussed in~\cite{pt2}.

  For events with open charm production, $PGF$ enters alone at leading
order (in the limit where the nucleon's intrinsic  charm is neglected). The main
difficulty lies in the combinatorial background associated with their
selection by the identification of a $D^o$ meson from its $K\pi$ decay.
This is a major concern in COMPASS where the vertex resolution is
not sufficient to reconstruct the decay vertex, because of the thickness
of the target. Special care is therefore taken to optimize the use of the data.
First, the favorable cases when the $D^o$ comes from a
$D^*\rightarrow D^o\pi$ decay are counted separately, {\it cf.} Fig.~\ref{DG}.
Secondly, a weighting procedure is applied for the
 derivation of
$\Delta G/G$:
  \[\Delta G/G = \frac{1}{P_T P_b f}~~\frac{\sum_i^{\Uparrow\uparrow} w_i - \sum_i^{\Uparrow\downarrow} w_i}{\sum_i^{\Uparrow\uparrow} w_i^2 + \sum_i^{\Uparrow\downarrow} w_i^2}~~~~~~~~~~w_i = \frac{\langle a_{LL}\rangle_i}{(1+B/S)_i}~~~~~~~~~~~~~~~~~~~~\]
where $a_{LL}$ is $PGF$'s analysing power and $S$ and $B$ are the signal and
background counting rates. In these conditions, the precision is expected to be $\delta\Delta G/G \simeq 0.24$, with direct and $D^*$-tagged $D^o$'s 
contributing about equally. This projection is plotted on Fig.~\ref{DG}.

\section{INCLUSIVE \boldmath$A_1$}
  The inclusive $A_1$ is derived from (\ref{AmuN}) through $DA_1 \simeq A_{{\mu}N}$,
where $D$ is the depolarization factor from the muon to the virtual photon.
  The data selection is actually not literally inclusive: at low $x$
a charged hadron is required in the final state, as was done by the
SMC experiment~\cite{A1}. This removes the
background from elastic and quasi-elastic scattering and allows for the
interaction vertex to be unambiguously located in either of the target
cells. The kinematical domain is restricted to $Q^2>1$ and
$0.1<y<0.9$, which makes the measurement cover nearly the same $x$ domain
as SMC.

  The results for the 2002-2003 data~\cite{publi} are shown in Fig.~\ref{A1}, together with results from SMC~\cite{A1}, HERMES~\cite{hermesA1}
and E155~\cite{e155}. They significantly improve the accuracy in the region $x < 0.03$.

\section{SEMI-INCLUSIVE \boldmath$A_1$}
  At the difference with the inclusive case, semi-inclusive measurements
allow to separate quark and anti-quark contributions. This is
possible when two conditions are met. First, the hard scattering and soft fragmentation
processes must be proved to factorize. Secondly, the current fragmentation
(arising
from the struck quark) must be clearly separated from the target one. 
One  also needs a large enough set of independent measurements in order to
get a meaningful flavor separation. COMPASS intends to compute the following
    asymmetries:
$
  A_1, A_1^{h^+},A_1^{h^-},A_1^{K^+},A_1^{K^-},A_1^{K^o_s}
  $, which, on its deuterium (isoscalar) target, gives access to the
 following combinations:
  $\Delta u+\Delta d,\Delta \bar{u} +\Delta \bar{d},\Delta s +
  \Delta \bar{s} $.

\begin{figure}[t]
  \begin{minipage}[b]{0.49\linewidth}
\includegraphics*[width=0.99\linewidth]{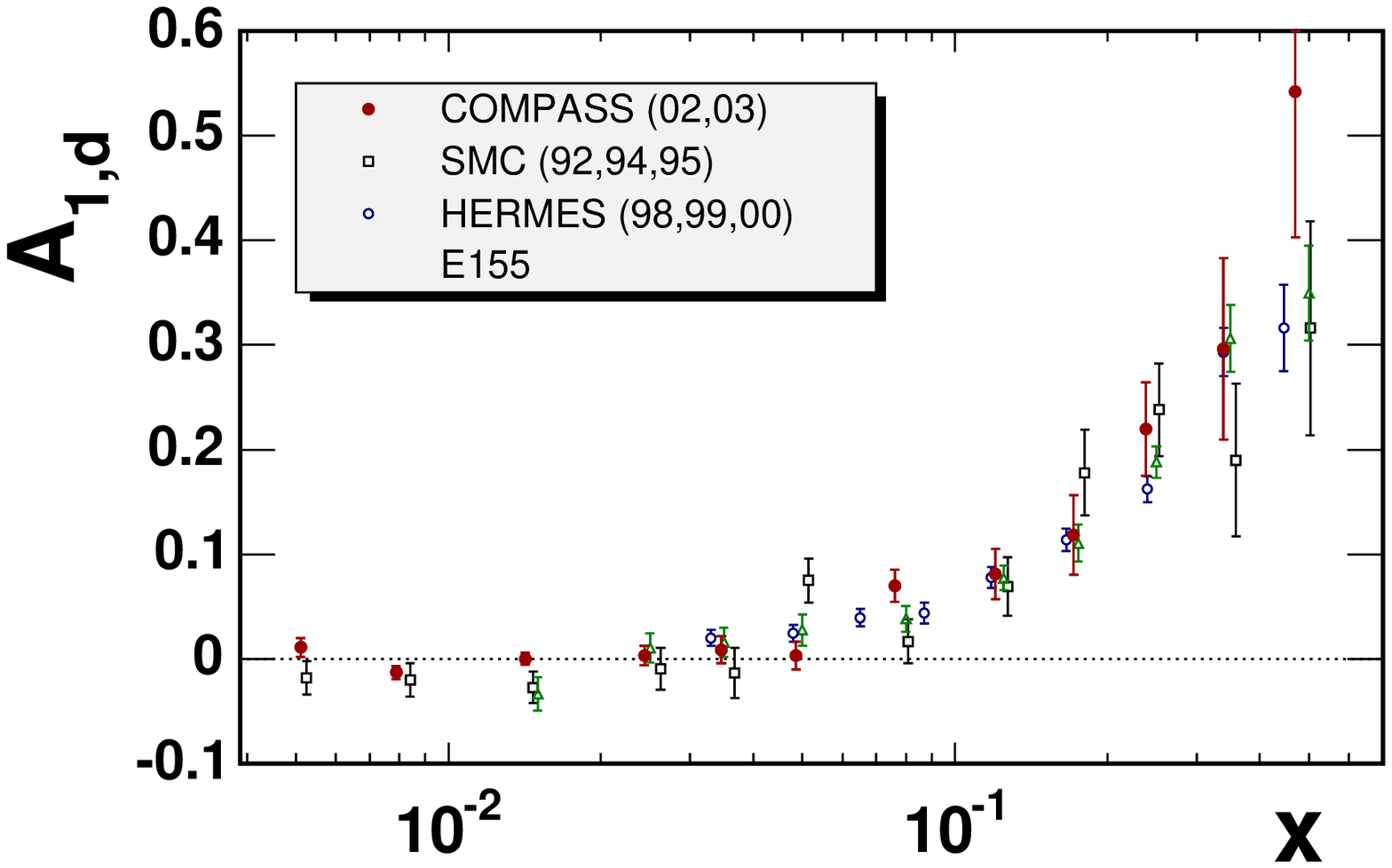}
  \end{minipage}
  \begin{minipage}[b]{0.49\linewidth}
\includegraphics*[width=0.99\linewidth]{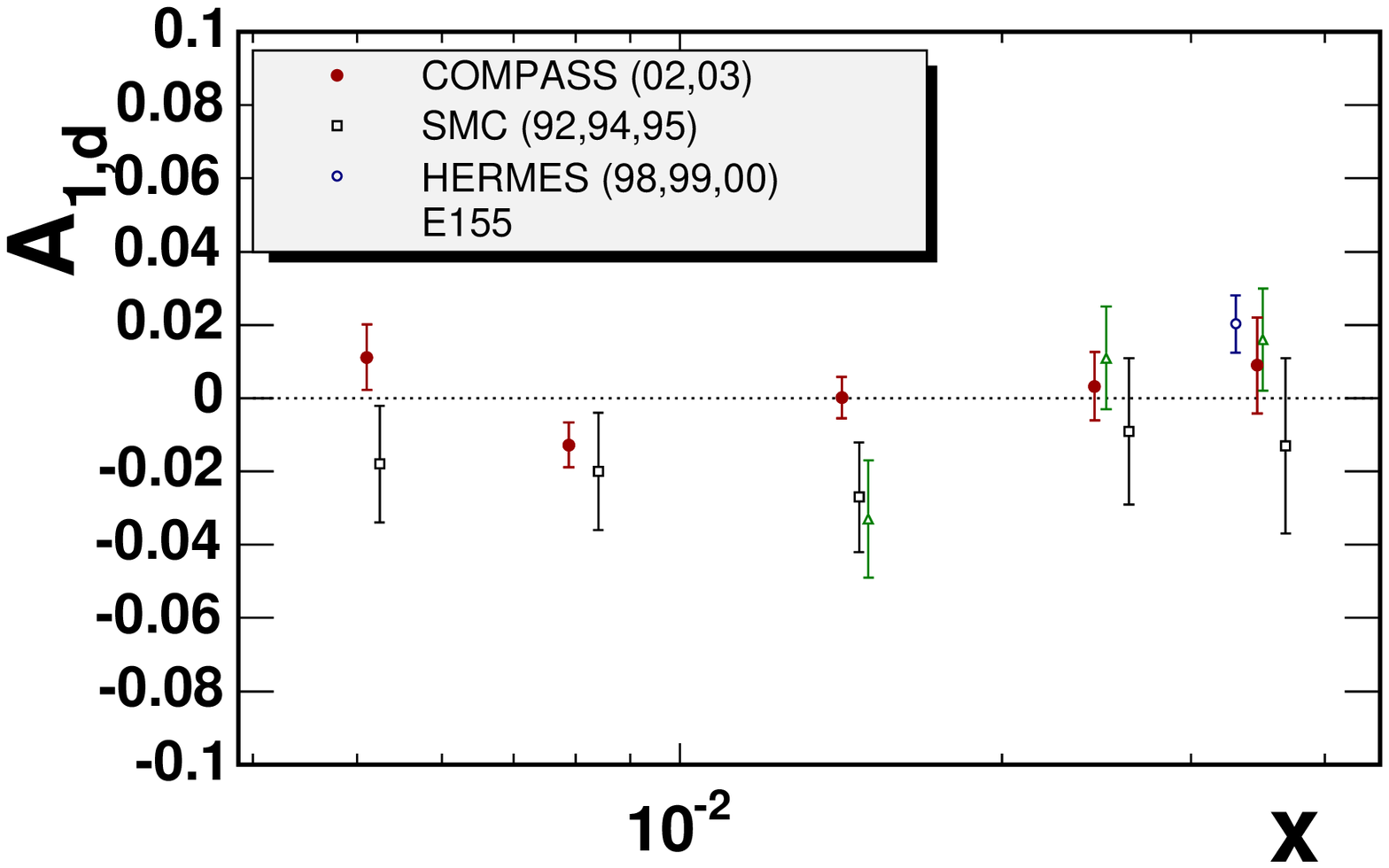}
  \end{minipage}
  \begin{minipage}[b]{0.49\linewidth}
\includegraphics*[width=0.99\textwidth]{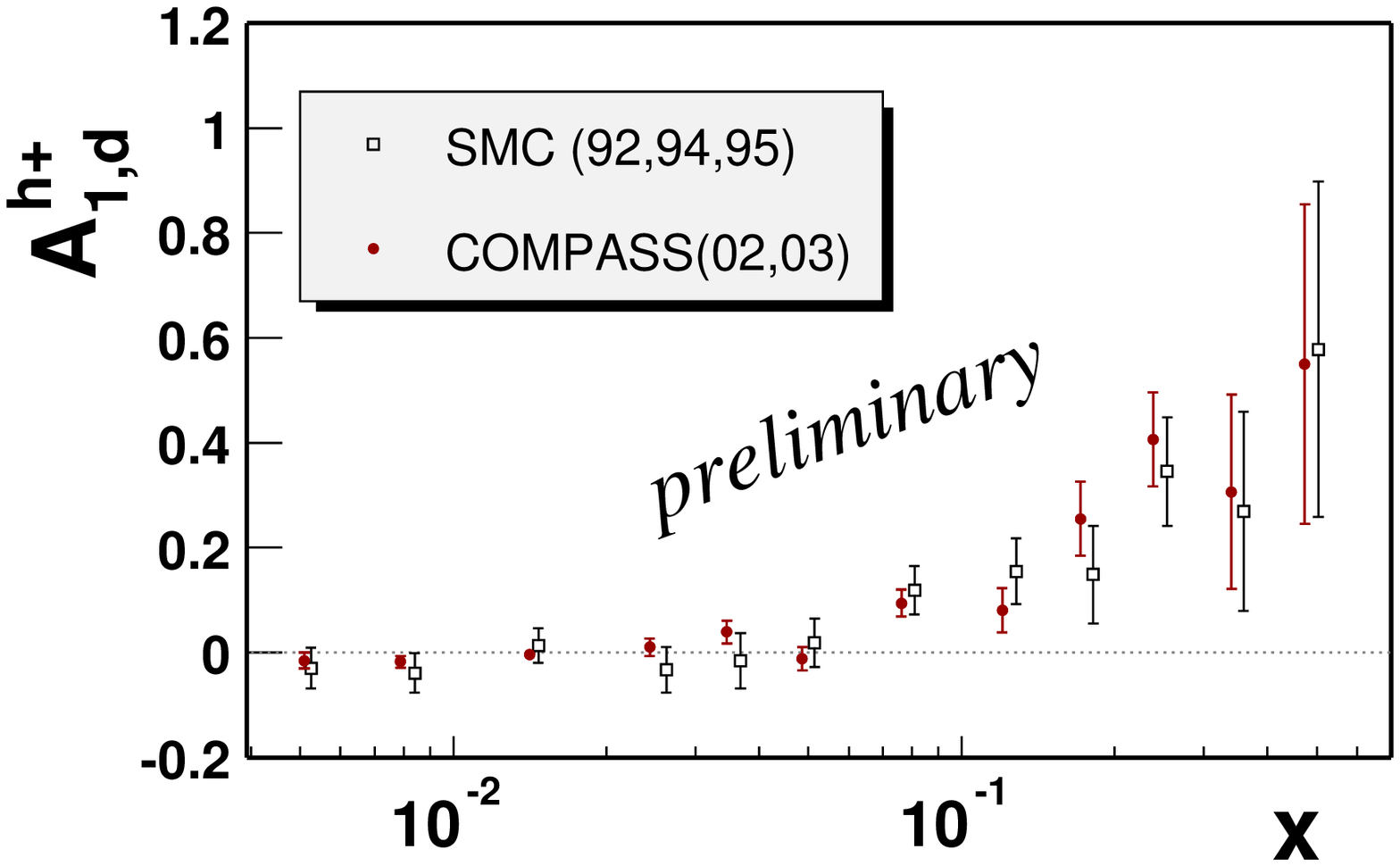}
  \end{minipage}
  \begin{minipage}[b]{0.49\linewidth}
\includegraphics*[width=0.99\textwidth]{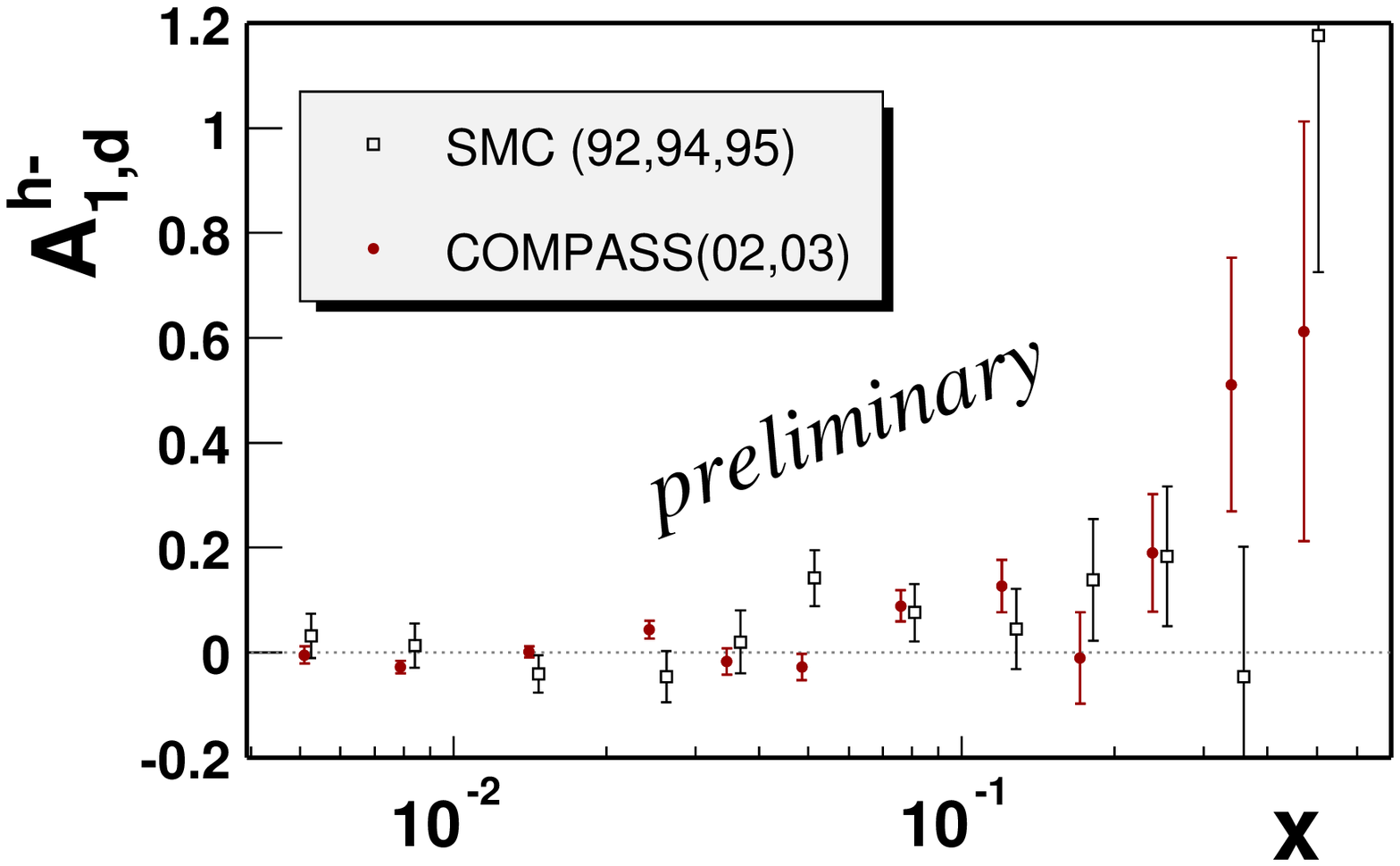}
  \end{minipage}
\vspace*{-1.cm}
\caption{$A_{1,d}~vs.~x$ from COMPASS 2002-2003 data compared with the results from other experiments ({\it see text for the references}). Top: Inclusive (with a zoom on the low $x$ region). Bottom: Semi-inclusive ($z=E_h/E_{\gamma^*}>0.2$ to select current fragmentation).}
\label{A1}
\end{figure}

  The analysis is under way. Preliminary results have been obtained
for the asymmetries $A_1^{h^+}$ and $A_1^{h^-}$  and are shown on Fig.~\ref{A1},
together with results from SMC~\cite{A1h}.

\section{OUTLOOK}
  COMPASS has demonstrated its ability to significantly contribute to the
understanding of the nucleon's spin. It is to complete the analysis of
its first 3 years of data by the end of 2005, yielding 2 measurements
of $\Delta G/G$, via the high $p_T$ and open charm channels, with respective
precisions of $\sim$0.05 and $\sim$0.24. While less precise, the open charm
channel will provide a model-free determination. The experiment will resume data taking in 2006.

\end{document}